\documentclass[preprint,showpacs,preprintnumbers,amsmath,amssymb,showkeys]{revtex4}

\usepackage{graphicx}% Include figure files
\usepackage{dcolumn}% Align table columns on decimal point
\usepackage{bm}% bold math

\begin{document}
%\setlength{\textheight}{7.7truein}  %for 2nd page o
%\def\bea{\begin{eqnarray}}
%\def\eea{\end{eqnarray}}
%\def\nn{\nonumber}

%\draft
%\preprint{}

\title{Quark-Lepton Similarity }

%\thispagestyle{empty}
%\begin{titlepage}

\author{Seungsu Hwang} \email{seungsu@cau.ac.kr}
\affiliation{Department of Mathematics, Chung-Ang University,
Seoul 156-756, Korea}
\author{Kim Siyeon} \email{siyeon@cau.ac.kr}
\affiliation{Department of Physics,
        Chung-Ang University, Seoul 156-756, Korea}

\date{August 19, 2010}
\begin{abstract}
We propose the lepton mixing matrix at high energy scale to be
connected to quark mixing matrix by the similar transformation. The similarity between CKM and PMNS
significantly narrows down the ranges in physical parameters. The condition requires $\sin\theta_{13}$ not to be larger than 0.15,
masses to be of quasi-degenerate normal ordering, and $\tan\beta$ to be large.
\end{abstract}

\pacs{11.30.Fs, 14.60.Pq, 14.60.St}

%\end{titlepage}
\maketitle \thispagestyle{empty}

%%%%%%%%%%%%%%%%%%%%%%%%%%%%%%%%%%%%%%%%%%%%%%%%%%%%%%%%%%%%

\section{Introduction}

The theory of grand unification (GUT) ties quarks and leptons in a
representation. When a unification group, e.g., SO(10), $E_6\,$ or
others, broke down to the group of Standard Model, quarks and
leptons find their own bases, though they were once in a common flavor
basis. In the light of the philosophy of unification, it is
natural that the separated quark sector and lepton sector do
appear with common properties in mass and mixing. Apparently,
however, the low energy phenomenology does not provide a clue of a
common basis shared by them. The
Cabibbo-Kobayashi-Maskawa(CKM) matrix of quark mixing, to be
denoted by $U_Q$, and the Potecorvo-Maki-Nakagawa-maskawa(PMNS)
matrix of lepton mixing, to be denoted by $U_L$, can be expressed
commonly in standard parametrization;
    \begin{footnotesize}
    \begin{eqnarray}
        \label{standard}
        \left(\begin{array}{ccc}
            c_{12}c_{13} & s_{12}c_{13} & s_{13}e^{-i\delta}\\
            -s_{12}c_{23}-c_{12}s_{23}s_{13}e^{i\delta} &
            c_{12}c_{23}-s_{12}s_{23}s_{13}e^{i\delta} &
            s_{23}c_{13} \\
            s_{12}s_{23}-c_{12}c_{23}s_{13}e^{i\delta} &
            -c_{12}s_{23}-s_{12}c_{23}s_{13}e^{i\delta} &
            c_{23}c_{13}
        \end{array}\right)
    \end{eqnarray}
    \end{footnotesize}
\noindent where $s_{ij}$ and $c_{ij}$ denote $\sin{\theta_{ij}}$
and $\cos{\theta_{ij}}$, respectively, of a mixing angle
$\theta_{ij}$ between $i$-th and $j$-th generations.
The global fit of CKM matrix in the Standard Model implies the best-fit values, $s_{12}^Q=0.2276, ~s_{23}^Q=0.04148,
~s_{13}^Q=0.00359$, and $\delta=1.20$ rad, while the best-fit values of
lepton mixing angles are $s_{12}^L=0.551, ~s_{23}^L=0.707,$ and
$s_{13}^L=0.1$\cite{Amsler:2008zzb}\cite{GonzalezGarcia:2007ib}. Quarks and leptons do barely share common
properties from the low-energy phenomenological point of view.

Upon incomplete understanding of neutrino masses, reliable data is only the mass squared difference, so that the best-fit values of them, $\Delta m^2_{21}=7.7\times 10^{-5}\mathrm{eV}^2, ~|\Delta
m^2_{31}|=2.4\times 10^{-3}\mathrm{eV}^2,$ are the usable numbers allowed now\cite{GonzalezGarcia:2007ib}. Such partial information currently hold the wide
possibility for a change in PMNS matrix at high energy scale, since the masses can be either quasi-degenerate or hierarchical. It
is known that the quasi-degenerate type of mass spectrum helps
Renormalization Group Equations(RGE) drive the running of mixing
angles making a rapid ascent or descent, although hierarchical masses make the running of angles slow, or even unchanging
\cite{Chankowski:1999xc}\cite{Casas:1999ac}\cite{Antusch:2003kp}\cite{Dighe:2006sr}. That is, RG evolution of mixing angles based on the current data has ruled out neither the form of PMNS in high energy scale exactly equal to CKM nor the form of PMNS completely different from the CKM. Any type of connection between those extreme cases, of course, has not been excluded.

Quark Lepton Unification(QLU) implies that quarks and leptons have the common flavor basis and the common mass basis so that CKM and PMNS have the same elements
\cite{Mohapatra:2003tw}\cite{Hosteins:2006ja}\cite{Agarwalla:2006dj}. Here is introduction to the simplest extension to QLU that might be brought from the basis shift occurred during the seesaw mechanism or extra symmetry breaking process: The PMNS is a similar transformation of CKM, which is possible when the unitary transformation from old flavor basis to new flavor basis is equal to the one from old mass basis to new mass basis. The condition will be called Quark Lepton Similarity(QLS). The special case in which the unitary matrix for the similar transformation is an identity matrix corresponds to QLU. The prediction from the similarity condition narrows down the
ranges in a number of physical parameters such as  $\Delta m^2_{31}$, $\sin\theta_{13}\sin\delta$, and $\tan\beta$. This paper is organized as follows. Sec.II describes the way to express the similarity condition of CKM and PMNS in terms of lepton mixing angles, and Sec.III deals with the results of RGE of lepton mixing
angles and their comparison with the angles drawn up from the
similarity. In conclusion, the prediction from QLS will be
summarized.

\section{Quark Lepton Similarity}

If mixing matrices are built on flavor basis and mass basis within a GUT symmetry
that strongly unifies quarks and leptons, it is naturally proposed
that there is one mixing matrix at the GUT scale $M_G$\cite{Mohapatra:2003tw}\cite{Agarwalla:2006dj};
     \begin{eqnarray}
     |~m_{Q,L}~\rangle=U_G~|~Q,L~\rangle,
     \end{eqnarray}
or $|~m_Q~\rangle=U_G~|~Q~\rangle$ and $|~m_L~\rangle=U_G~|~L~\rangle$.  However it is
also possible to have separated bases after the GUT symmetry
breaking, either quarks or leptons, to be transformed from the
original common basis to a new one. Seesaw mechanism or extra
symmetry breaking might be a process to bring such basis shift.
So, if the lepton basis is the one under change, the mixing matrix $U_L$ is defined for a new flavor basis and a new mass basis such that $|~m_L'~\rangle=U_L~|~L'~\rangle$.
As long as the total transition probability is conserved,
there exists a unitary transformation between $|~L~\rangle$ and $|~L'~\rangle$ and another between $|~m_L~\rangle$ and $|~m_L'~\rangle$, denoting them by $P$ and $P'$, respectively. Our assumption $P=P'$
gives rise to a connection between the original mixing matrix $U_G$
and the lepton mixing matrix $U_L$, such as $U_G = P^{-1} U_L P$. When $U_Q=U_G$ or even when $U_Q$ undergoes another basis shift as $U_L$ did, one can obtain $U_Q = P^{-1} U_L P$.
Then it is formally said that the two matrices $U_Q$ and $U_L$ are
similar, or unitarily equivalent for $P^{-1}=P^\dagger$.

Similar matrices have the same characteristic polynomials,
    \begin{eqnarray}
        p(\lambda)=\det U -c_2 \lambda + tr\,U \lambda^2 - \lambda^3,
        \label{character}
    \end{eqnarray}
for an eigenvalue $\lambda$ of matrix $U$. Since both $U_Q$ and $U_L$ are unitary, their equal determinants are trivial; $\det U_Q=\det U_L=1$. The principal minor $c_2$ of the matrix in Eq.(\ref{character}) is the trace of the minor matrix $\widetilde{U}$. An element $\widetilde{U}_{ij}$ is given by the determinant of the $2\times 2$ matrix with row $i$ and column $j$ removed in $U$. The comparison of $c_2$'s of two matrices and the comparison of their traces will fulfill the matching of the characteristic polynomials. Hereafter we use vanishing functions $W$ and $\widetilde{W}$ defined with
similar matrices $U_Q$ and $U_L$;
    \begin{eqnarray}
        W(\theta_{12}, \theta_{23}, \theta_{13}, \delta)=tr\, U_Q - tr\,U_L,  \label{similar}
    \end{eqnarray}
while $\widetilde{W}(\theta_{12}, \theta_{23}, \theta_{13}, \delta)=tr\, \widetilde{U}_Q - tr\,\widetilde{U}_L,$ with
lepton mixing angles and phase in $U_L$; $\theta_{12}, \theta_{23}, \theta_{13}$ and $\delta$. According to
Eq.(\ref{standard}), the trace of $U_L$ is
    \begin{eqnarray}
    tr\, U_L
    =c_{12}c_{13}+ c_{23}c_{13}+c_{12}c_{23}
    -s_{12}s_{23}s_{13}e^{i\delta},
    \end{eqnarray}
whereas $ tr\, \widetilde{U}_L = c_{12}c_{13} + c_{23}c_{13} + c_{12}c_{23} -s_{12}s_{23}s_{13}e^{-i\delta}$. Separating real parts and imaginary parts, $\widetilde{W}=0$ is equivalent to $W=0$, which is sufficient to show the similarity between $U_Q$ and $U_L$.

\begin{figure}
\resizebox{70mm}{!} %\resizebox{150mm}{!}
{\includegraphics[width=0.75\textwidth]{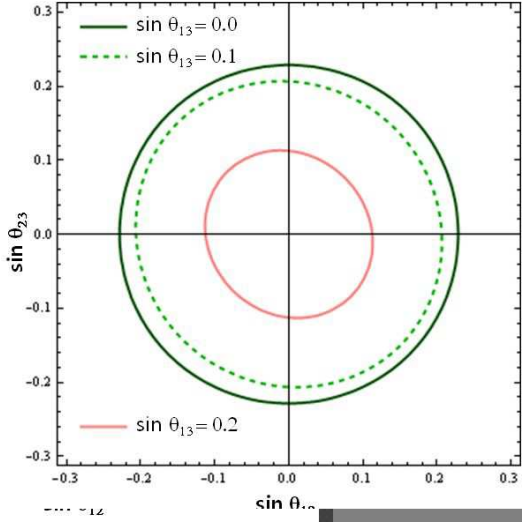}}
% Here is how to import EPS art
\caption{\label{fig:similar}$W(\theta_{12}, \theta_{23})=0$.}
\end{figure}

The similarity between CKM and PMNS could be considered at the
Seesaw Mechanism scale $M_R$, which might be GUT scale $M_G$
or somewhere several order of magnitude below $M_G$. It has been
checked that the elements of the CKM matrix do not vary over
scales by renormalization group equations due to strong mass
hierarchy. The trace of $U_Q$ obtained with the best-fit values
of angles, i.e., $tr\,U_Q$, is $2.947-i(3.13\times10^{-5})$. Then, the real part of $W$ in
Eq.(\ref{similar}) reduces to
    \begin{eqnarray}
    && W_r = 2.95 - c_{12}c_{13} - c_{23}c_{13} - c_{12}c_{23}
    + s_{12}s_{23}s_{13}\cos\delta, \nonumber \\
    &&
    \delta=\arcsin(\frac{3.13\times10^{-5}}{s_{12}s_{23}s_{13}}),
    \label{REsimilarity}
    \end{eqnarray}
where the expression $\delta$ is obtained from the imaginary parts
of the traces. The condition does not require the specification of $\delta$. The similarity condition $W=0$ is adopted to find the
lepton mixing angles $\theta_{12}, \theta_{23}$, and $\theta_{13}$
at $M_R$ scale. Fig.\ref{fig:similar} shows loci of the possible
pairs of $\theta_{12}, \theta_{23}$ that satisfy the similarity
for some fixed values of $\theta_{13}$.

\section{Renormalization Group Equations for neutrino mixing angles}

The renormalization group equations (RGE) of mixing angles are
well summarized in Refs.\cite{Casas:1999ac}\cite{Mohapatra:2003tw} so that they can be
rearranged such as
    \begin{eqnarray}
        && \dot{s}_{23} = -H_\tau c_{23}^2
            ( -s_{12}U_{\tau1}A_{31}+c_{12}U_{\tau2}A_{32}),
            \label{dsdt23} \\
        && \dot{s}_{13} = -H_\tau c_{23} c_{13}^2
            ( c_{12}U_{\tau1}A_{31}+s_{12}U_{\tau2}A_{32}),
            \label{dsdt13} \\
        && \dot{s}_{12} =
            -H_\tau c_{12}(c_{23}c_{13}^2s_{12}U_{\tau1}A_{31}
            \nonumber \\
        && \hspace{27pt}-c_{23}s_{13}c_{12}U_{\tau2}A_{32}
            + U_{\tau1}U_{\tau2}A_{21}),
    \label{dsdt12}
    \end{eqnarray}
where $U_{\alpha i}$ is an element of the matrix in
Eq.(\ref{standard}) and $A_{ij}=(m_i+m_j)/(m_i-m_j)$. According to Ref.\cite{Agarwalla:2006dj}, since $\frac{d\delta}{dt}$ for zero Majorana phases is proportional to $\sin\delta$, the top-down RG solution of $\delta$ approaches zero as $\Lambda$ goes to $M_Z$. The RGE of $\delta$ might be skipped without affecting physical implication.
The RGE of
masses to run together with Eq.(\ref{dsdt23})~-~Eq.(\ref{dsdt12}) is
    \begin{eqnarray}
        \dot{m_i}=-m_i(2H_\tau U_{\tau1}^2-H_u).
    \end{eqnarray}
For the SM
    \begin{eqnarray}
        && H_\tau=\frac{3}{32\pi^2}(h^-_\tau)^2 \label{tau_minus}\\
        && H_u=\frac{1}{16\pi^2}
        \{3g_2^2-2\lambda-6(h^-_t)^2-6(h^-_b)^2-2(h^-_\tau)^2\},
        \nonumber
    \end{eqnarray}
while for the MSSM
    \begin{eqnarray}
        && H_\tau=-\frac{1}{16\pi^2}(h^+_\tau)^2 \label{tau_plus} \\
        && H_u=\frac{1}{16\pi^2}\{\frac{6}{5}g_1^2+6g_2^2-6(h^+_t)^2\},
        \nonumber
    \end{eqnarray}
where the Yukawa couplings satisfy the boundary conditions at the
SUSY threshold, $M_S$, at which are
$h^+_t(M_S)=h^-_t(M_S)/\sin\beta$ and
$h^+_{b,\tau}(M_S)=h^-_{b,\tau}(M_S)/\cos\beta$
\cite{Barger:1992ac}\cite{Anderson:1992ba}. The accompanied
RGE in one-loop order of the gauge couplings, the Yukawa
couplings, and the Higgs coupling are taken from the Ref.\cite{Barger:1992ac}.
As for the boundary condition, the Yukawa couplings $h_b$ and
$h_\tau$ are unified at $M_G=10^{16} GeV$. The unification of
Yukawa couplings as well as gauge couplings pass a threshold at
$M_S=1 TeV$ switching the gauge frame from the SM to MSSM. The
chosen seesaw scale is $M_R=10^{11} GeV$.

\begin{figure}
\resizebox{70mm}{!}{\includegraphics[width=0.75\textwidth]{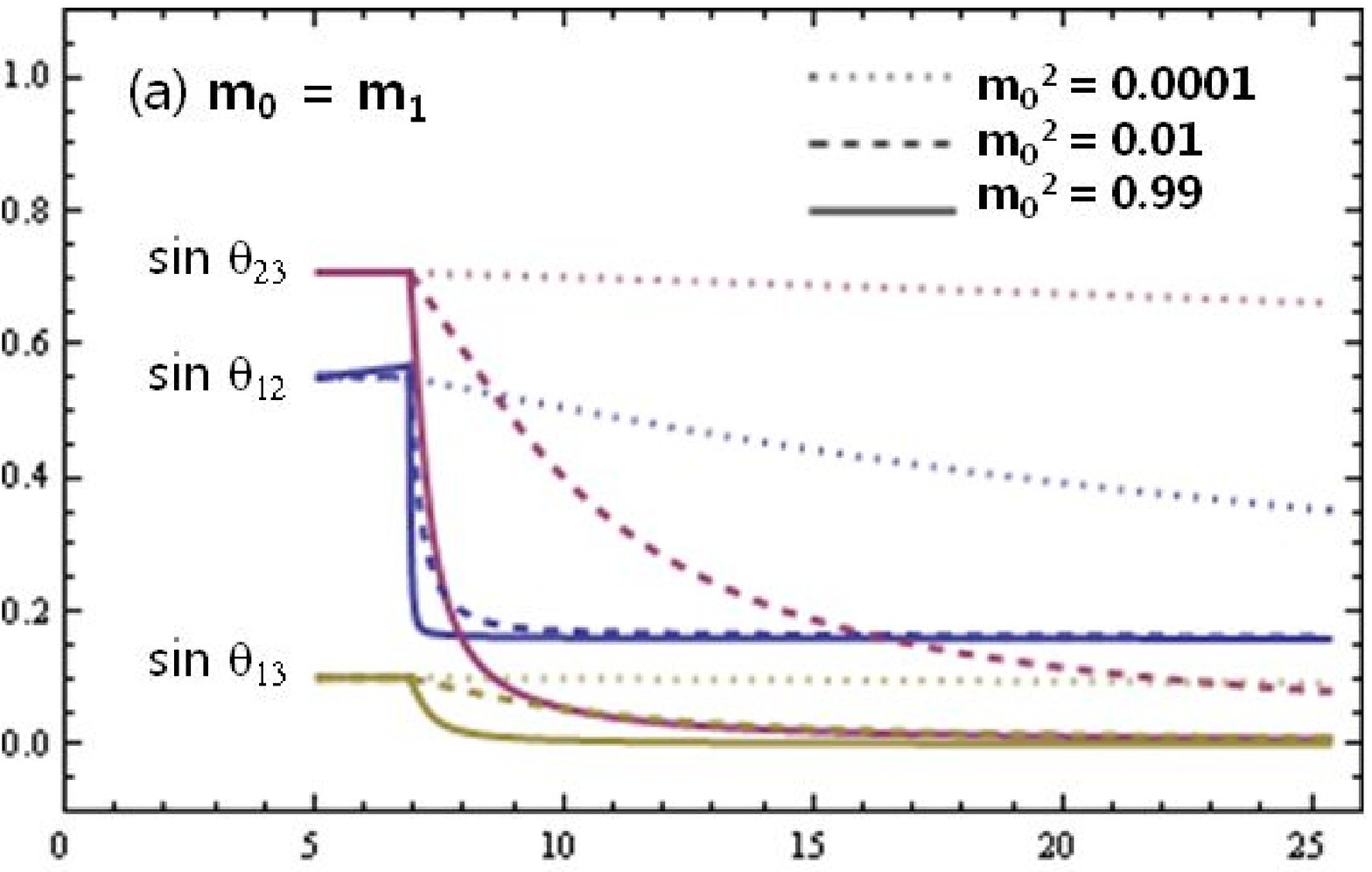}} \\
\resizebox{70mm}{!}{\includegraphics[width=0.75\textwidth]{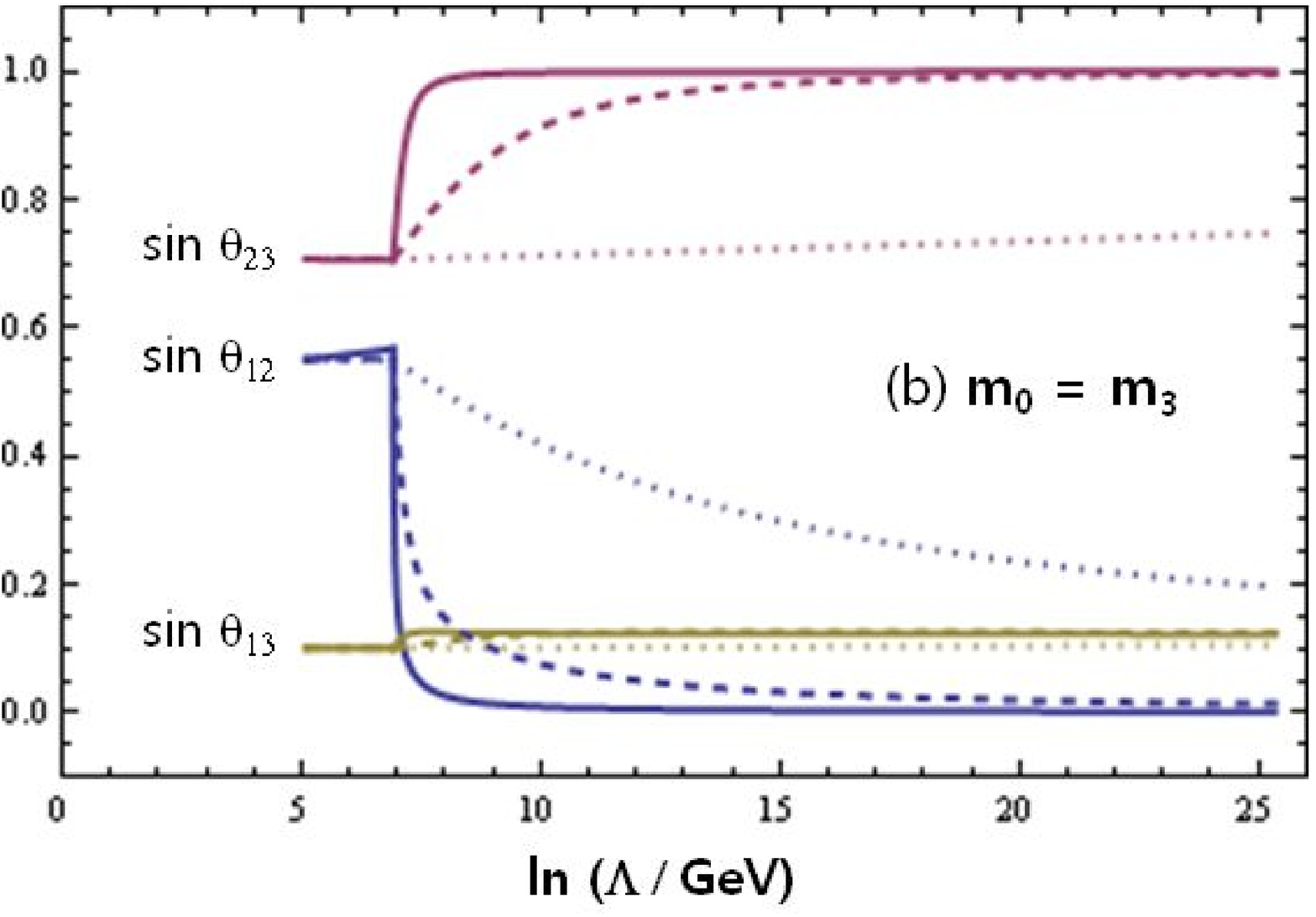}}
\caption{\label{fig:rge_msqr}  Evolution of $\sin\theta_{23}$,
$\sin\theta_{12}$, and $\sin\theta_{13}$ for $\tan\beta=50$.}
\end{figure}

\begin{figure}
\resizebox{70mm}{!}{\includegraphics[width=0.75\textwidth]{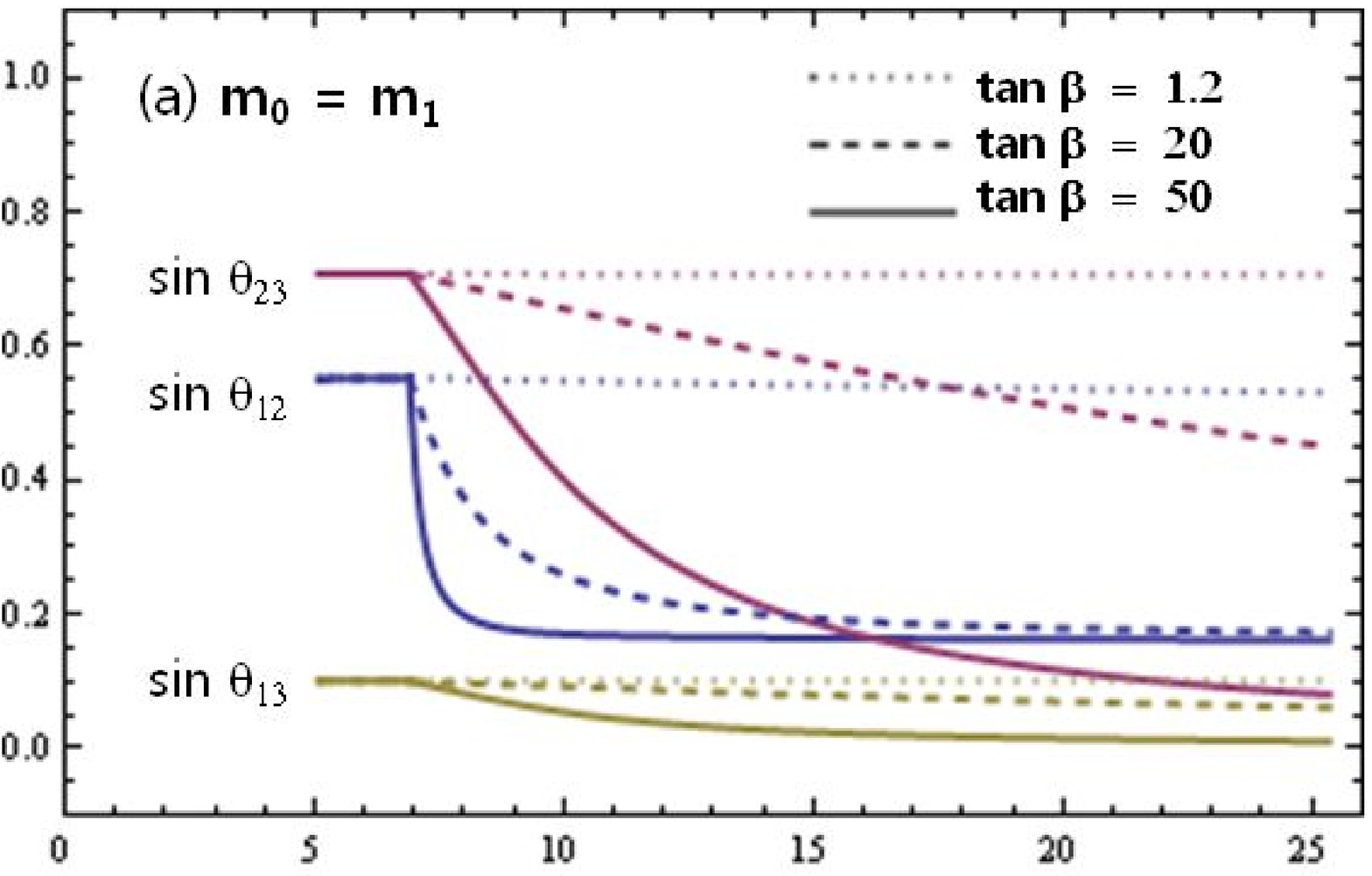}} \\
\resizebox{70mm}{!}{\includegraphics[width=0.75\textwidth]{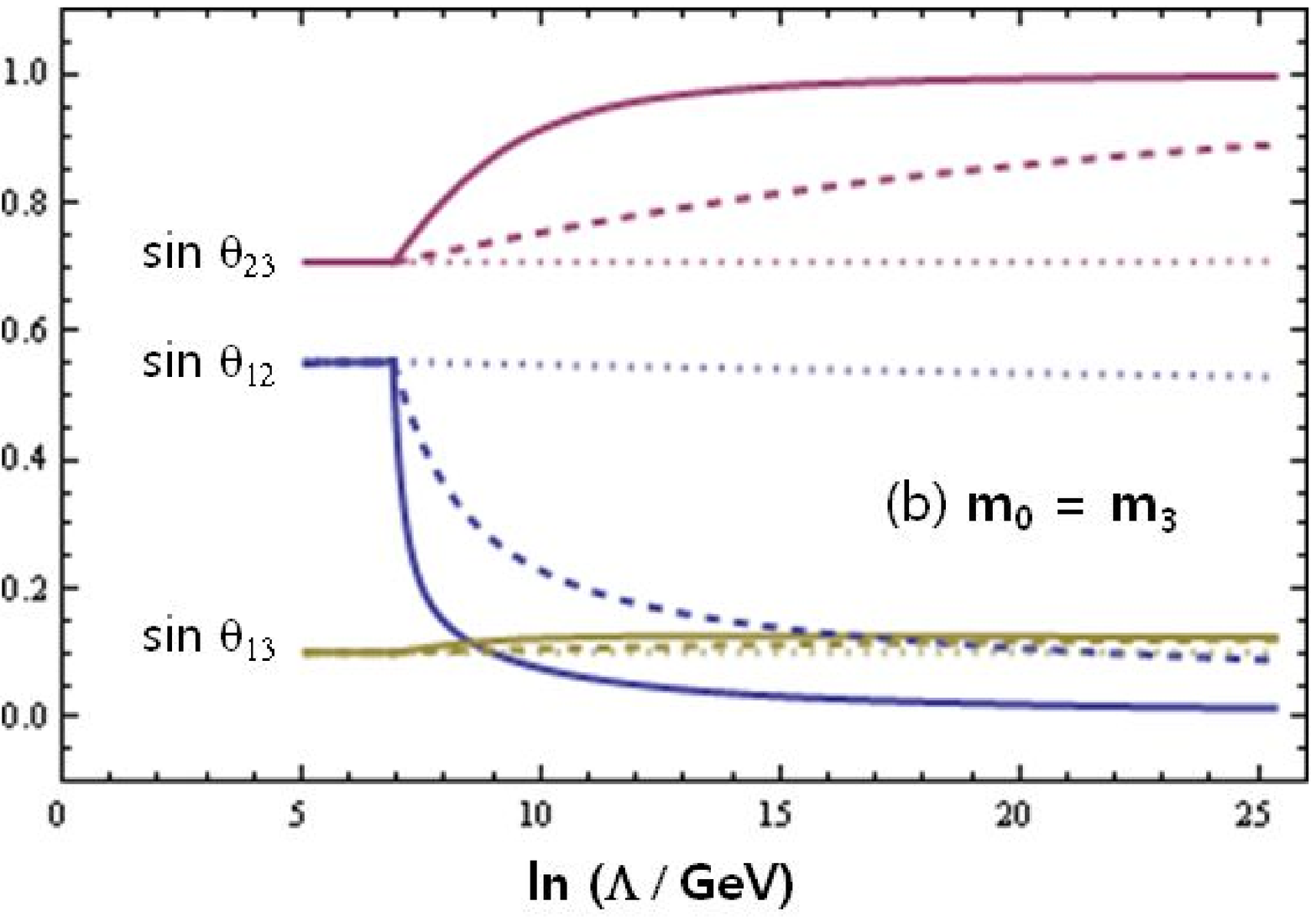}}
\caption{\label{fig:rge_beta}  Evolution of $\sin\theta_{23}$,
$\sin\theta_{12}$, and $\sin\theta_{13}$ for $m_0^2=0.1$.}
\end{figure}

The running of the mixing angles over scales is very subtle to the
type of mass spectrum due to the factor $A_{ij}$ as shown in
Eq.(\ref{dsdt23})~-~Eq.(\ref{dsdt12}). The current experimental
data indicate neutrino mass spectrum to be one of the following
two cases. First, normal ordering, $m_1<m_2<m_3$, is one candidate, which can be given by
    \begin{eqnarray}
        && m_1^2 = m_0^2, \nonumber \\
        && m_2^2 = m_0^2 + \Delta m^2_{21}, \label{msqr_nh}\\
        && m_3^2 = m_0^2 + |\Delta m^2_{32}| \nonumber,
    \end{eqnarray}
where $m_0$ denotes a parameter meaning the lightest mass, and
$\Delta m^2_{21}$ and $|\Delta m^2_{32}|$ are simply the dictation
to the results of the solar and atmospheric neutrino oscillation.
The other is inverted ordering, $m_3<m_1<m_2$, which can be given by $m_3^2 = m_0^2,~ m_1^2 = m_0^2 + |\Delta m^2_{32}| - \Delta m^2_{21},$ and $m_2^2 = m_0^2 + |\Delta m^2_{32}|.$
Cases of the normal ordering and the inverted ordering will be represented by $m_0=m_1$ and $m_0=m_3$, respectively, and their effects on RG running of the angles are separated into (a) and (b) in Fig.\ref{fig:rge_msqr} and Fig.\ref{fig:rge_beta}.

The strong hierarchy in neutrino masses appears with a small value of $m_0^2$ so that the value of $m_3^2$ in Eq.(\ref{msqr_nh}) could be in order of $|\Delta m^2_{32}|$, while the quasi-degeneracy in masses appears with a large value of $m_0^2$ so that the value of $m_3^2$ in Eq.(\ref{msqr_nh}) could be in order of $m_0^2$. The initial conditions in the bottom-up integrations consist of $\sin\theta^Z_{23}=0.707, ~\sin\theta^Z_{12}=0.551$, and $\sin\theta^Z_{13}=0.10$. The upper index Z is used to denote the value evaluated at $M_Z$. The curves in Fig.\ref{fig:rge_msqr} explain the running aspects for three types of masses; strong hierarchy, weak hierarchy and quasi-degeneracy. The evolutions in angles for quasi-degeneracy are quite rapid, which reflect the direct contribution of $A_{ij}$ in Eq.(\ref{dsdt23})~-~Eq.(\ref{dsdt12}). In comparison, the masses in strong hierarchy do not help the mixing angles vary as energy scale goes up. In Fig.\ref{fig:rge_beta}, the influences of $\tan\beta$ on the evolution of the angles are featured. The changes in curves for large $\tan\beta$ reveal the contribution of $H_\tau$ in Eq.(\ref{dsdt23})~-~Eq.(\ref{dsdt12}). The Yukawa coupling $h_\tau$ in $H_\tau$ in Eq.(\ref{tau_minus}) is
magnified by the factor $(\cos\beta)^{-1}$ at the threshold
$M_S$, as shown in Eq.(\ref{tau_plus}). Then the curves of
$\sin\theta$'s above the threshold become drastically rapid for a
large $\tan\beta$, as appeared in Fig.\ref{fig:rge_beta}.

\begin{figure*}
        \resizebox{150mm}{!}
        {\includegraphics[width=0.75\textwidth]{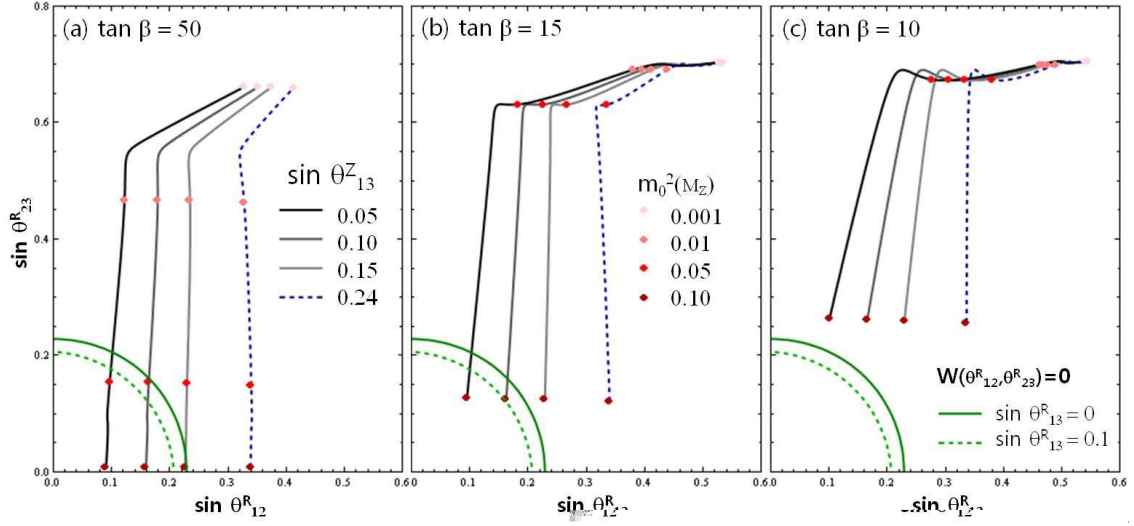}}
        \caption{\label{fig:tanbeta}
        The solutions of RGE at $M_R$, $(\sin\theta^R_{12},\sin\theta^R_{23})$, featuring the dependency on $\tan\beta$ with (a),(b) and (c), the dependency on $\sin\theta^Z_{13}$ with various gray lines, and the dependency on $m_0^2$ along with the curves.}
    \end{figure*}

\begin{figure*}
        \resizebox{150mm}{!}
        {\includegraphics[width=0.75\textwidth]{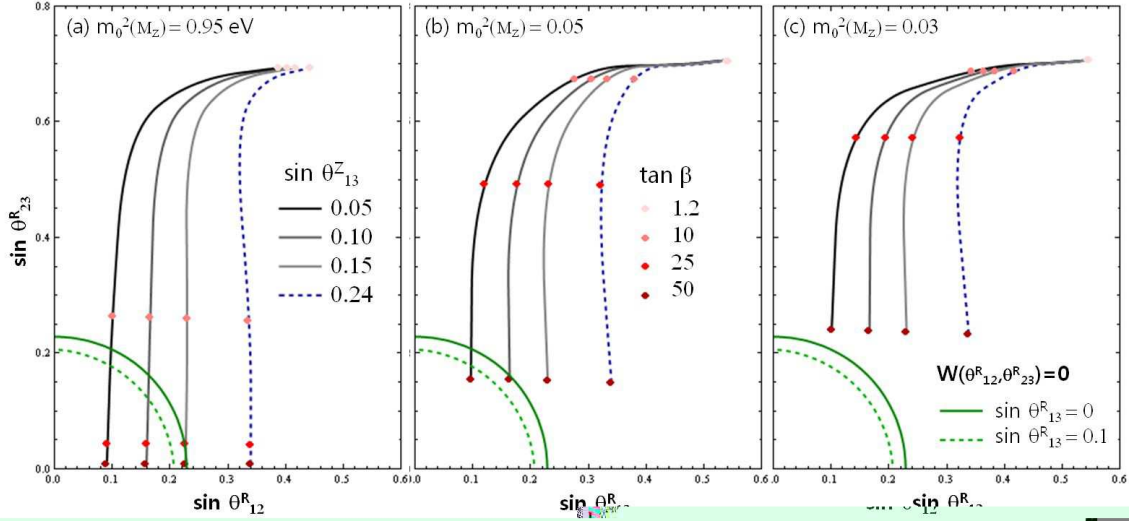}}
        \caption{\label{fig:msqr}
        The solutions of RGE at $M_R$, $(\sin\theta^R_{12},\sin\theta^R_{23})$, featuring the dependency on $m_0^2$ with (a),(b) and (c), the dependency on $\sin\theta^Z_{13}$ with various gray lines, and the dependency on $\tan\beta$ along with the curves.}
    \end{figure*}

Fig.\ref{fig:tanbeta} and
Fig.\ref{fig:msqr} show the solutions of RGE in
Eq.(\ref{dsdt23})~-~Eq.(\ref{dsdt12}) at $M_R$, $(\sin\theta_{12}^R, \sin\theta_{23}^R)$, which will be compared with the angles obligated to the quark-lepton similarity. The upper index R is used to denote the value evaluated at $M_R$. Fig.\ref{fig:tanbeta} describes their
continuous dependency on $m_0^2$ at $M_Z$ with respect to given
values of $\tan\beta$ and $\sin\theta^Z_{13}$, while Fig.\ref{fig:msqr}
describes their continuous dependency on $\tan\beta$ with respect
to given values of $m_0^2$ and $\sin\theta^Z_{13}$. The different brightness in gray curves is originated from a different value in $\sin\theta^Z_{13}$. 
The first quadrant of $W(\theta_{12},\theta_{23})=0$ in
Fig.\ref{fig:similar} is substituted in Fig.\ref{fig:tanbeta} and
Fig.\ref{fig:msqr}, and so its intersection with a curve will find
the angles to make $U_L$ similar to $U_Q$. The plots clearly indicate that only the solution obtained with large $\tan\beta$ and large $m_0$ can make the intersection. On the other hand, the solutions with small $\tan\beta$ and small $m_0$ converge to $(\sin\theta_{12}^Z, \sin\theta_{23}^Z)$.

From
Fig.\ref{fig:rge_msqr} and Fig.\ref{fig:rge_beta}, it is worth
keeping in mind that $\theta^R_{13}\lesssim\theta^Z_{13}$ for
$m_0=m_1$. Thus, the values of
$\theta^R_{13}$ on the gray curves near the similarity shell are all ranged within 0 to 0.1. In fact, they are almost close to zero, since the evolution of $\sin\theta_{13}$ with large $\tan\beta$ and quasi-degenerate masses approaches zero as the scale goes up. In other words, the solution of $\theta^R_{13}$ in RGE can find its matching point to the similarity just inside the shell drawn in Fig.\ref{fig:tanbeta} and
Fig.\ref{fig:msqr}. Although here the value of $\theta^R_{13}$ is not specified, 
the intersection with the shell between 
$W|_{\sin\theta^R_{13}=0}=0$ and $W|_{\sin\theta^R_{13}=0.1}=0$ can guarantee the solution to the similarity
$W(\theta^R_{12}, \theta^R_{23})=0$, technically.
    
For inverted mass ordering, as mentioned in Ref.\cite{Dighe:2006sr}, the RG running
effect in mixing angles except for $\theta_{12}$ appears as rising as the scale $\Lambda$ increases, i.e., $\theta^R_{13}\gtrsim\theta^Z_{13}$ and $\theta^R_{23}\gtrsim\theta^Z_{23}$. That is opposite to the
running effects in $\theta_{23}$ and $\theta_{13}$ for normal
mass ordering. The running effect for $m_0=m_3$ case is described in
Fig.\ref{fig:rge_msqr}(b) and Fig.\ref{fig:rge_beta}(b). The increasing
curve in $\sin\theta_{23}$ from the initial value
$\sin\theta_{23}(M_Z)=\pi/4$ as $\Lambda$ increases makes the
result of RGE rather away from the $W=0$ shell. Thus, there is no
solution to the similarity that is compatible with the mass type
of $m_0=m_3$, whether it is hierarchical or quasi-degenerate. Last, even in the most optimistic case with $\tan\beta$ and masses, the initial condition $\sin\theta^Z_{13} > 0.15$ cannot result in the evolution of angles within the $W=0$.

\section{Discussion on phenomenological implication}

The Standard Model and its extension for massive neutrinos can be
considered as an effective theory stemmed from a fundamental
theory such as GUT. If a mechanism like Seesaw Mechanism makes the
basis of leptons different from that of quarks pointing the
significant distinction between them in masses and mixing angles,
there could exist the similarity between CKM and PMNS, which means
that one is the similar transformation of the other. The allowed
ranges in lepton mixing angles are strongly restricted by the
similarity constraint. The similarity was applied at $M_R$, and
the angles at the scale were evaluated as results of RGE under
various combination of the initial conditions with $m_0,
\tan\beta$, and $\sin\theta^Z_{13}$.

The physical implication from the quark lepton similarity is
summarized as follows: First, the inverse ordered mass type of
neutrino, either hierarchical or quasi-degenerate, is ruled out.
Second, only normal ordered quasi-degenerate mass type of neutrino
can satisfy the similarity, while the normal hierarchy mass type
does not. Third, a value of $\sin\theta^Z_{13}$ larger than 0.15
is ruled out, and a small value is preferred. Last, small
$\tan\beta$ cannot result in the RG effect to be compatible to the
similarity. Thus, the prediction from the model will be tested by the
current experiments to look for SUSY signals like
LHC and the various types of neutrino oscillation experiments. Especially, the transition probability in super-beam neutrino oscillation may take advantage of the capability to predict $s_{13}\cos\delta$ or $s_{13}\sin\delta$ as in Eq.(\ref{REsimilarity}) , not to predict simply $s_{13}$ or $\delta$, since that can reduce the ambiguity, so-called degeneracy problem, caused from multiple parameters\cite{Arafune:1997hd}\cite{Hwang:2007na}.

\begin{acknowledgments}
K.S. thanks members in the Physics Department at Cheonnam National
University for warm hospitality, where a part of this work was
completed. The authors thank P.S.Lee for help for figures and thank T.Lee, P.Ko and S.Y.Choi for helpful discussion. This research
was supported by Chung-Ang University research grants in 2008.
\end{acknowledgments}

\end{document}